\documentclass[runningheads]{llncs}
\usepackage{graphicx}
\usepackage{url} 
\usepackage{amsmath}
\usepackage{tabularx}
\usepackage{hhline}
\usepackage{tabularx}
\usepackage{graphicx}
\usepackage{amsmath}
\usepackage{amssymb}
\usepackage{url}
\usepackage{threeparttable}
\usepackage{framed}
\usepackage{multirow}
\usepackage{gensymb}
\usepackage{hhline}
\usepackage{rotating}
\usepackage{babel}
\usepackage{mathtools}
\usepackage{booktabs}
\usepackage{blindtext}
\usepackage{tikz}
\usepackage{graphicx,calc}
\newlength\myheight
\newlength\mydepth
\settototalheight\myheight{Xygp}
\newcommand*\inlinegraphics[1]{%
  \settototalheight\myheight{Xygp}%
  \settodepth\mydepth{Xygp}%
  \raisebox{-\mydepth}{\includegraphics[height=\myheight]{#1}}%
}
\usepackage[title]{appendix}
\begin{document}
\sloppy

\title{Anatomy Completor: A Multi-class Completion Framework for 3D Anatomy Reconstruction}
\titlerunning{3D Multi-class Anatomy Completor}
\author{Jianning Li\inst{1} \and
Antonio Pepe \inst{2} \and
Gijs Luijten \inst{1,2}\and
Christina Schwarz-Gsaxner\inst{2}\and
Jens Kleesiek\inst{1}\and
Jan Egger\inst{1,2}
}

\authorrunning{J. Li et al.}

\institute{Institute for AI in Medicine (IKIM), University Hospital Essen  \and Institute of Computer Graphics and Vision (ICG), Graz University of Technology }

\maketitle              % typeset the header of the contribution

\begin{abstract}
In this paper, we introduce a completion framework to reconstruct the geometric shapes of various anatomies, including organs, vessels and muscles. Our work targets a scenario where one or multiple anatomies are missing in the imaging data due to surgical, pathological or traumatic factors, or simply because these anatomies are not covered by image acquisition. Automatic reconstruction of the missing anatomies benefits many applications, such as organ 3D bio-printing, whole-body segmentation, animation realism, paleoradiology and forensic imaging. We propose two paradigms based on a 3D denoising auto-encoder (DAE) to solve the anatomy reconstruction problem: (i) the DAE learns a \textit{many-to-one} mapping between incomplete and complete instances; (ii) the DAE learns directly a \textit{one-to-one} residual mapping between the incomplete instances and the target anatomies. We apply a loss aggregation scheme that enables the DAE to learn the \textit{many-to-one} mapping more effectively and further enhances the learning of the residual mapping. On top of this, we extend the DAE to a multiclass completor by assigning a unique label to each anatomy involved. We evaluate our method using a CT dataset with whole-body segmentations. Results show that our method produces reasonable anatomy reconstructions given instances with different levels of incompleteness (i.e., one or multiple random anatomies are missing). Codes and pretrained models are publicly available at \url{https://github.com/Jianningli/medshapenet-feedback/tree/main/anatomy-completor}.

\end{abstract}

\section{Introduction}
3D anatomy reconstructions play important roles in medical applications and beyond, such as (1) 3D bio-printing and organ transplantation, where damaged/diseased organs from traumatic injuries or pathologies are replaced by 3D bio-printed artificial organs \cite{parihar20213d}; (2) paleoradiology and forensic imaging, in which the full anatomical structures are re-established based on the skeleton remains \cite{wilkinson2010facial,la20223d,missal2023forensic}; (3) whole-body segmentation, where pseudo labels of whole-body anatomies are generated given only sparse manual annotations \cite{jaus2023towards,seibold2023accurate,wasserthal2022totalsegmentator}; (4) animation realism \cite{ali2013anatomy}; and (5) diminished reality, where the 3D view of an anatomy blocked by medical instruments is reconstructed. Such an anatomy reconstruction task is well aligned with the shape completion problem in computer vision, which is commonly solved based on the symmetry of geometric shapes \cite{shi2022learning} or using learning-based approaches, where auto-encoder and generative adversarial networks (GANs)  \cite{yan2022shapeformer,chibane2020implicit,wodzinski2022deep,Sarmad_2019_CVPR} are popular choices. Recent years have witnessed a growing interest in medical shape completion, with the rapid development of medical deep learning \cite{egger2022medical}. Nevertheless, existing works in this direction are mostly focused on reconstructing a pre-defined and geometrically simple bone structure, such as the cranium \cite{kodym2020skull,kodym2021deep,li2021autoimplant,li2021automatic,li2023towards,wodzinski2022deep,matzkin2020self,morais2019automated}, maxilla \cite{zhang2020cleft}, spine \cite{meng2020learning} and teeth \cite{toscano2023teeth}, which restricts their scope of application to implant and prosthetic design. Existing methods for medical shape completion are commonly based on variants of auto-encoder and U-Net \cite{kodym2020skull} and statistical shape models (SSMs) \cite{goparaju2022benchmarking,pimentel2020automated}. Reconstructing random anatomies with varied geometric complexity is significantly harder than when the reconstruction target is pre-defined as in prior works. To realize the former, a network learns not only to identify the targets (i.e., what are missing in the input) but to reconstruct them, a process analogous to object instance segmentation \cite{He_2017_ICCV}, where a network first identifies all objects in an image and then segments them. However, random anatomy reconstruction has not been covered by existing research, which only completes one fixed anatomy with missing part(s), and remains to be an open problem. The goal of this work is to extend medical shape completion to the whole body, covering the majority of anatomy classes, and to realize random anatomy reconstruction in a single shape completion framework. To achieve this goal, we derived a 3D anatomical shape dataset from a fully-segmented CT dataset and trained a 3D convolutional denoising auto-encoder on the dataset to learn a mapping relationship between the incomplete instances and the corresponding targets, i.e., the full segmentations or the missing anatomies. Both quantitative and qualitative evaluations have demonstrated the effectiveness of our proposed method towards solving the anatomical shape reconstruction problem. 

\begin{figure}
\centering
\includegraphics[width=\linewidth]{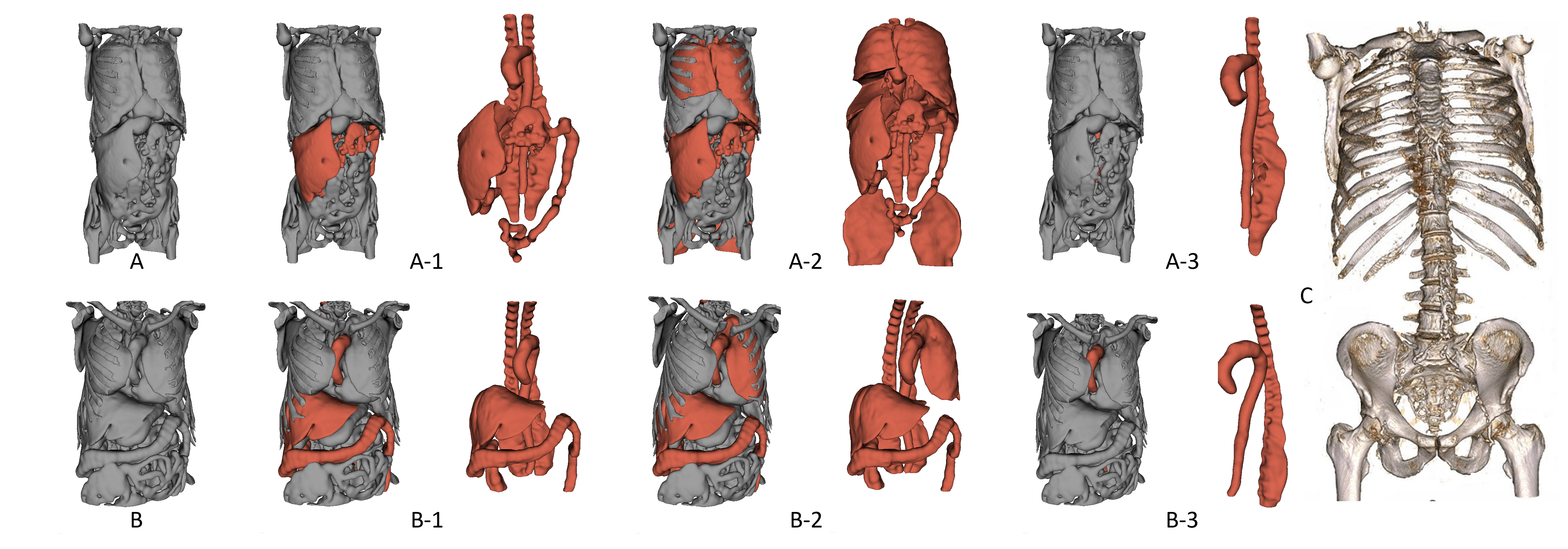}
 \caption{Illustration of the pre-processed dataset. (A, B): the full anatomy segmentations from two subjects. (A-1, A-2, A-3) and (B-1, B-2, B-3): three incomplete instances with random missing anatomies (shown in red). (C): the skeleton in a CT scan.}
\label{dataset}
\end{figure}

\section{Methods}

\subsection{Problem Formulation}
\label{problem_formulation}
Reconstructing random missing anatomies is formulated as a shape completion problem, where the goal is to learn a mapping $\mathcal{F}$ between the incomplete instances from $N$ subjects $\mathcal{X}=\begin{Bmatrix}x^m_{n}\end{Bmatrix}^{m=1,...,M}_{n=1,...,N}$ and the corresponding complete ground truth $\mathcal{Y}=\begin{Bmatrix}y_n\end{Bmatrix}^N_{n=1}$ derived from whole-body anatomy segmentations. For subject $x_n$, there exist $M$ instances i.e., $x^1_n, x^2_n, ..., x^m_n, ..., x^M_n $ with different degrees of incompleteness, where one or multiple random anatomies are missing. Therefore, $\mathcal{F}$ is supposed to be a \textit{many-to-one} mapping, i.e.,

\begin{equation}
\mathcal{F}: \begin{Bmatrix}x^m_n \end{Bmatrix}^M_{m=1}\rightarrow y_n, \, n=1,2,...,N
\label{mapping}
\end{equation}

\noindent We use binary voxel grids to represent 3D anatomies, such that $x^m_n,y_n \in R^{L\times W \times H}$. The value of a voxel in $x^m_n$, $y_n$ is '1' if the voxel belongs to an anatomy and '0' otherwise. Such a formulation extends existing medical shape completion methods that target only a single, pre-defined anatomy to random anatomies.

\subsection{Denoising Auto-encoder with Residual Connections}

Given the notations in Section \ref{problem_formulation}, the missing anatomies for subject $x_n$ can be conveniently expressed in a residual form: $\begin{Bmatrix}
y_n-x^m_n\end{Bmatrix}^M_{m=1}$. Therefore, apart from learning the full mapping $\mathcal{F}$, we can instead learn a residual mapping $\mathcal{F}_{res}$:

\begin{equation}
\mathcal{F}_{res}: \begin{Bmatrix}x^m_n \end{Bmatrix}^M_{m=1}\rightarrow \begin{Bmatrix}
y_n-x^m_n\end{Bmatrix}^M_{m=1}, \, n=1,2,...,N
\label{residual_mapping}
\end{equation}

\noindent Unlike $\mathcal{F}$, the residual mapping $\mathcal{F}_{res}$  is obviously \textit{one-to-one}, which can be straightforwardly realized based on deep residual learning \cite{he2016deep}. Motivated by this observation, we propose to solve the shape completion problem using a 3D denoising auto-encoder (DAE) with a residual connection between the input and the output. The input $x^m_n$ is treated as a corrupted version of $y_n$ with random noise. The DAE denoises the input by restoring the anatomies missing in $x^m_n$. The DAE is trained in a supervised fashion, with the input being $\mathcal{X}$ and the ground truth being $\mathcal{Y}$.  Even though both mappings are learnable by the DAE, we presume that a \textit{one-to-one} mapping relationship is easier to learn than a \textit{many-to-one} mapping, so that the DAE can reach a superior reconstructive performance by learning $\mathcal{F}_{res}$.

\subsection{Loss Aggregation for Random Anatomy Completion}
To learn the \textit{many-to-one} mapping $\mathcal{F}$, we train the DAE by optimizing a Dice loss function $\mathcal{L}_{dice}$ aggregated over $M$ versions of incomplete instances with random missing anatomies:

\begin{equation}
  \mathcal{L}_{\mathcal{F}}= \sum_{m=1}^{M}\sum_{n=1}^{N}\mathcal{L}_{dice}(y_n,\tilde{y}^m_n)
  \label{agg_dice_f}
\end{equation}

\noindent where $\mathcal{L}_{dice}=\frac{2\sum (y_n\bigodot\hat{y}^m_n)}{\sum (y_n\bigodot y_n)+\sum (\hat{y}^m_n\bigodot \hat{y}^m_n)}$ is the standard Dice loss \cite{milletari2016v}. $\hat{y}^m_n$ denotes the prediction for $x^m_n$ given the mapping $\mathcal{F}$, and $\bigodot$ denotes the Hadamard product (i.e., element-wise multiplication between two matrices). $\sum$ denotes the summation of all the elements of a matrix. Optimizing such an aggregated loss function $\mathcal{L}_{\mathcal{F}}$ ensures that the DAE learns to reconstruct a complete set of anatomies regardless of the class and$/$or number of anatomies that are absent in the input. Similarly, to learn the \textit{one-to-one} residual mapping $\mathcal{F}_{res}$, the following loss function is optimized:
\begin{equation}
  \mathcal{L}_{\mathcal{F}_{res}}=\sum_{m=1}^{M} \sum_{n=1}^{N}\mathcal{L}_{dice}(y_n,\tilde{x}^m_n+x^m_n)
  \label{agg_dice_fres}
\end{equation}

\noindent where $\tilde{x}^m_n$ denotes the reconstructed missing anatomies for $x_n$. Depending on the mapping to be learned, the respective loss function ($\mathcal{L}_{\mathcal{F}}$ or $\mathcal{L}_{\mathcal{F}_{res}}$) is used.

\subsection{Multi-class Anatomy Completion}
For the multi-anatomy completion task, compared to representing $x^m_n$ and $y^n$ as binary voxel grids in which different anatomies are not differentiated (Section 2.1), it is more desirable to assign a unique label to each anatomy in $x^m_n$ and $y_n$. This extension can be easily achieved by setting the number of output channels of the penultimate layer of the DAE network to the number of anatomy classes. Each channel predicts the probability of occupancy of the voxel grids for an anatomy. The same Dice loss $\mathcal{L}_{dice}$ can be calculated between the output and the ground truth in one-hot encoding.

\section{Experiments and Results}
\subsection{Dataset and Pre-processing}
We validate our method using a public CT dataset with whole-body anatomy segmentations, which is publicly available at \url{https://zenodo.org/record/6802614#.Y_YMwXbMIQ8}. The dataset comprises 1024 CT images, each accompanied by a set of segmentation masks of 104 anatomies (organs, bones, muscles, vessels) \cite{wasserthal2022totalsegmentator}. After screening (discarding images with corrupted segmentations), 737 sets of segmentations are included in this work, which are further randomly split into a training (451) and test set (286). For each set of segmentations, we randomly remove anatomies accounting for at least 10\%, 20\% and 40\% of the entire segmentation's volume to create the incomplete instances $\mathcal{X}$. The original segmentations serve as the ground truth $\mathcal{Y}$. Considering that anatomy ratios are subject-specific, different type and$/$or number of anatomies could have been removed for different subjects given the same threshold, as can be seen from Figure \ref{dataset}. Thus, anatomy removal is analogous to inserting random noise to $\mathcal{Y}$. In general, using a 10\% threshold (Figure \ref{dataset}, A-2, B-2) removes more anatomies than using higher thresholds (20\% and 40\%), and using a threshold of 40\% removes only large anatomies, such as the aorta and the autochthonous back muscles (Figure \ref{dataset}, A-3, B-3). The small bones such as the individual ribs and vertebrae that form the skeleton (Figure \ref{dataset}, C) enclosing the internal anatomies are generally not removed, providing a natural constraint for anatomy reconstruction. We use the ratio-based method to remove anatomies, so that each full segmentation yields three instances with random incompleteness in the training and test set. We denote the three test sets as $D_{test1}$ (10\%), $D_{test2}$ (20\%) and $D_{test3}$ (40\%). Besides random anatomy removal, we create another test set $D_{test4}$ by removing only one specific anatomy from the full segmentations randomly selected from the test set. All the images are re-scaled to a uniform size of $128^3$ ($L,W,H=128$).  We made the anatomical shape dataset used in this study publicly available through \textit{MedShapeNet} \cite{li2020baseline}.

\subsection{Implementation Details}
The DAE is comprised of four two-strided 3D convolutional (conv3D) and transposed convolutional (t\_conv3D) layers for downsampling and upsampling. To increase the learning capacity, we add a single-strided conv3D layer after each t\_conv3D layer, and further append four single-strided conv3D layers at the end of the DAE. We use \textit{ReLu} activations and a kernel size of three for all layers, amounting to around 22M trainable parameters. The residual connection is implemented as an addition between the input and the output of the penultimate layer. The DAE is implemented using TensorFlow \cite{abadi2015tensorflow} and trained on an NVIDIA RTX 3090 GPU using the ADAM optimizer \cite{kingma2014adam}. The learning rate is set to 0.0001 and the exponential decay rate for the first moment estimates is set to 0.3 for the ADAM optimizer.

\subsection{Experimental Setup}
Since, to our knowledge, our paper is the first to investigate random anatomy reconstruction, we adhere to the following steps to validate our methods: (i) A baseline is established by training the DAE without residual connection using a conventional Dice loss from existing single anatomy completion studies \cite{matzkin2020self,li2021autoimplant}; (ii) On top of the baseline, we train the DAE using the aggregated Dice loss (Equation \ref{agg_dice_f}); (iii) We train the DAE with residual connection (Equation \ref{residual_mapping}) using a conventional Dice loss; (iv) We train the DAE with residual connection using the aggregated Dice loss (Equation \ref{agg_dice_fres}). For all experiments, the DAE is trained for 100 epochs. The baseline experiment evaluates the feasibility of realizing random anatomy reconstruction using a single shape completion framework, and experiments (ii-iv) verify the effectiveness of each proposed components (i.e., residual connection, loss aggregation) for the anatomy reconstruction task. We denote the trained DAE models from experiment (i-iv) as $DAE_b$, $DAE_{agg}$, $DAE_{res}$ and $DAE_{agg+res}$, respectively. Dice similarity coefficient (DSC) is used for quantitative evaluation of the results on test set $D_{test1}$, $D_{test2}$, and $D_{test3}$. The output of the DAE is interpolated to the original size to calculate the DSC against the ground truth. On $D_{test4}$, we perform an empirical evaluation of our method in reconstructing one specific anatomy.

\subsection{Results}

\begin{table*}[ht]
\centering
\caption{Mean (Standard Deviation) of DSC on $D_{test1}$, $D_{test2}$, $D_{test3}$}
\begin{tabular}[t]{cccc} 
\toprule
Methods  & $D_{test1}$ & $D_{test2}$ & $D_{test3}$ \\
\hline
$DAE_b$&0.783 (0.075)&0.778 (0.061) & 0.757 (0.058)\\
$DAE_{agg}$&0.789 (0.073) & 0.803 (0.059)&0.812 (0.053)\\
$DAE_{res}$&\textbf{0.865} (0.069)&0.885 (0.046)&0.887 (0.047)\\
$DAE_{agg+res}$&\textbf{0.865} (0.074) & \textbf{0.904} (0.039)&\textbf{0.931} (0.030)\\
\bottomrule
\end{tabular}
\label{table:quantitative_results}
\end{table*}

\subsubsection{Quantitative Evaluation and Statistical Comparison} 
Table \ref{table:quantitative_results} presents the quantitative results of the ablation experiments, where the mean and standard deviations (SD) of DSC on test set $D_{test1}$, $D_{test2}$ and $D_{test3}$ are reported. The quantitative comparisons show that both loss aggregation ($DAE_{agg}$) and residual connection ($DAE_{res}$) help improve the anatomy reconstruction performance compared to the baseline ($DAE_b$). Furthermore, the comparison between $DAE_{agg}$ and $DAE_{res}$ demonstrates that the DAE is significantly better at learning the residual (Equation \ref{residual_mapping}) than the full anatomy (Equation \ref{mapping}). Combining both components ($DAE_{agg+res}$) further improves the reconstructive performance of the DAE compared to using each component individually. Furthermore, $DAE_{agg}$, $DAE_{res}$ and $DAE_{agg+res}$ also perform more stably across test instances (smaller SD) than the baseline, on all three test sets. Compared with $D_{test1}$ and $D_{test2}$, we notice an obvious drop of mean DSC on $D_{test3}$ for the baseline model, suggesting that $DAE_b$ tends to perform worse when the combined ratio of all missing anatomies becomes smaller. The combined ratio of all missing anatomies in $D_{test3}$ is likely to be lower, since fewer anatomies can be removed due to the higher ratio threshold. Higher sensitivity is required to detect and reconstruct smaller anatomies. Applying loss aggregation ($DAE_{agg}$) enforces the \textit{many-to-one} mapping and therefore mitigates the low sensitivity issue. The residual mapping ($DAE_{res}$) overcomes the low-sensitivity issue even without loss aggregation. A statistical comparison of the DSC between different models on the three test sets is also performed based on a t-test, and the $p$ values are reported in Table \ref{table:quantitative_results_stistical}. $p < 0.05$ indicates a statistically significant improvement. Based on Table \ref{table:quantitative_results} and the statistical comparisons of $DAE_{agg}\leftrightarrow DAE_b$ and $DAE_{agg+res}\leftrightarrow DAE_{res} $, we can also conclude that loss aggregation does not significantly improve the results on $D_{test1}$, which has a very high combined ratio of missing anatomies.

\begin{table*}[ht]
\centering
\caption{Statistical Comparison of DSC on Test Set $D_{test1}$, $D_{test2}$, $D_{test3}$ Between Different Methods. The Table Reports the $p$ Values From a T-test.}
\begin{tabular}[t]{cccc} 
\toprule
Methods  & $D_{test1}$ & $D_{test2}$ & $D_{test3}$ \\
\hline
$DAE_{agg}\leftrightarrow DAE_b$&0.328&4.301e-07 & 8.176e-29\\
$DAE_{res}\leftrightarrow DAE_b$ &2.839e-35 &2.147e-86 & 7.427e-114\\
$DAE_{agg+res}\leftrightarrow DAE_b$& 2.295e-33&1.437e-110&2.985e-163\\
$DAE_{res}\leftrightarrow DAE_{agg} $ &9.931e-32&3.051e-59&5.644e-57 \\
$DAE_{agg+res}\leftrightarrow DAE_{res}$ &0.989&2.866e-07&1.797e-34 \\
\bottomrule
\end{tabular}
\label{table:quantitative_results_stistical}
\end{table*}

\begin{figure}[h!]
\centering
\includegraphics[width=0.9\linewidth]{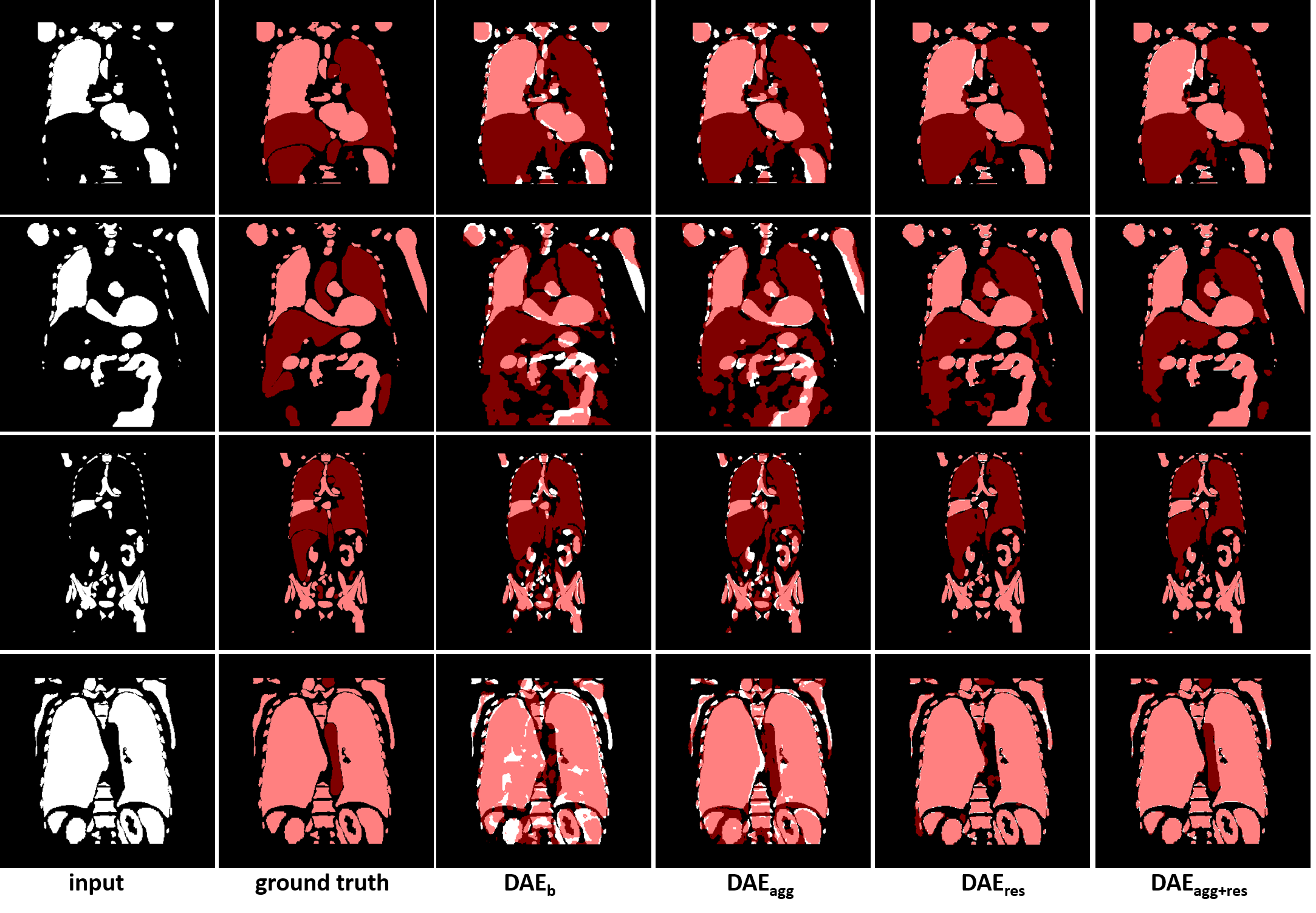} 
 \caption{Qualitative comparison of anatomy reconstruction performance. \protect\inlinegraphics{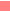} indicates the overlap between the reconstruction and the input, and \protect\inlinegraphics{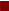} indicates the reconstructed missing anatomies. Small white blocks in the reconstructions indicate false negative predictions. }
\label{completion_results}
\end{figure}

\begin{figure}[h!]
\centering
\includegraphics[width=\linewidth]{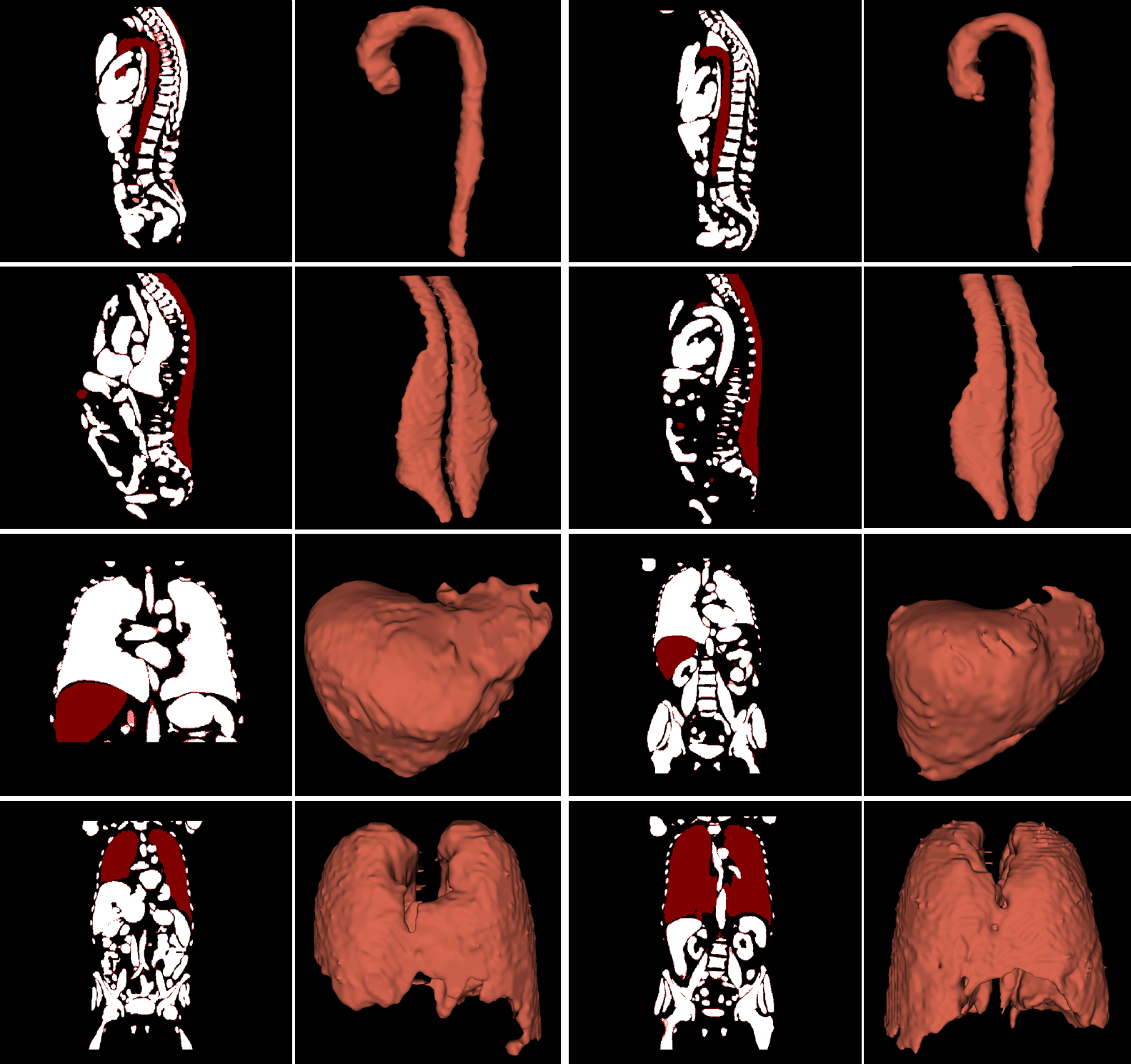}
 \caption{The first to last row show the reconstructed aorta, autochthonous back muscles, liver and lung by $DAE_{agg+res}$. Two test instances are presented for each anatomy class.}
\label{anatomy_results}
\end{figure}

\subsubsection{Qualitative Evaluation} Figure \ref{completion_results} illustrates the reconstruction results in 2D coronal planes. Multiple test instances with different degrees of incompleteness are presented. As seen from the ground truth  (Figure \ref{completion_results}, second column), an ideal reconstruction covers 100\% of the input and does not extend beyond the region enclosed by the ribs (Figure \ref{completion_results}, first column). The qualitative comparison shows that the DAE models trained for full anatomy reconstruction ($DAE_b$ and $DAE_{agg}$, Equation \ref{mapping}) have a tendency to produce false negatives, i.e., they fail to fully reconstruct existing anatomies, as shown by the small white blocks in the third and fourth column of Figure \ref{completion_results}, as well as false positives , i.e, they generate a reconstruction beyond the missing anatomies. Resorting to residual learning ($DAE_{res}$ and $DAE_{agg+res}$, Equation \ref{residual_mapping}) obviously mitigates the false prediction issue. Figure \ref{anatomy_results} shows the reconstruction results from the best performing model $DAE_{agg+res}$ for a single anatomy, specifically the aorta, the autochthonous back muscles, liver and lung. For single anatomy reconstruction, only one random anatomy is missing in the input ($D_{test4}$). For smaller anatomies like the kidney and spleen, these models are not sufficiently sensitive to detect their absence and produce a reasonable reconstruction.  Only for relatively large anatomies, such as livers and lungs, single anatomy reconstruction is feasible (Figure \ref{anatomy_results}). Increasing the loss aggregation scope (i.e., the $M$ in Equation \ref{agg_dice_f}, \ref{agg_dice_fres}) to explicitly cover the individual small anatomies during the training process is a promising solution to the low sensitivity problem. Appendix (A) provides preliminary results that support this observation regarding the reconstruction of small missing anatomies. In Appendix (B), we show that it is feasible to reconstruct the whole anatomies given only the skeleton (rib cage  + spine). These findings are potentially useful for (semi-)supervised whole-body segmentations, in which a human annotator provides manual segmentations for only a few of the anatomies, while the anatomy completor generates the segmentation masks in 3D for the rest. Even though the quality of the generated segmentations might not be sufficient to serve as the ground truth, they could be used as the initial pseudo labels that can be iteratively refined \cite{seibold2022reference}. Appendix (B) gives an extreme example where only the skeleton is given or annotated. It should be noted that the current results for such examples are not optimal, and serve only as a proof of concept.

\begin{figure}[h!]
\centering
\includegraphics[width=\linewidth]{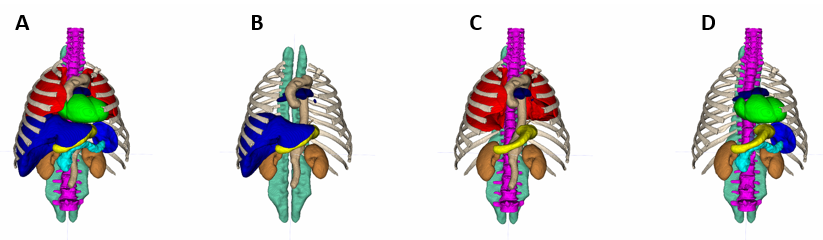}
 \caption{Dataset for the multi-class anatomy completor. (A) the 12 anatomy segmentations. (B-D) three incomplete instances where some of the 12 anatomies are missing.}
\label{multi_class}
\end{figure}

\subsubsection{Multi-class Anatomy Completion}
\begin{figure}[h!]
\centering
\includegraphics[width=\linewidth]{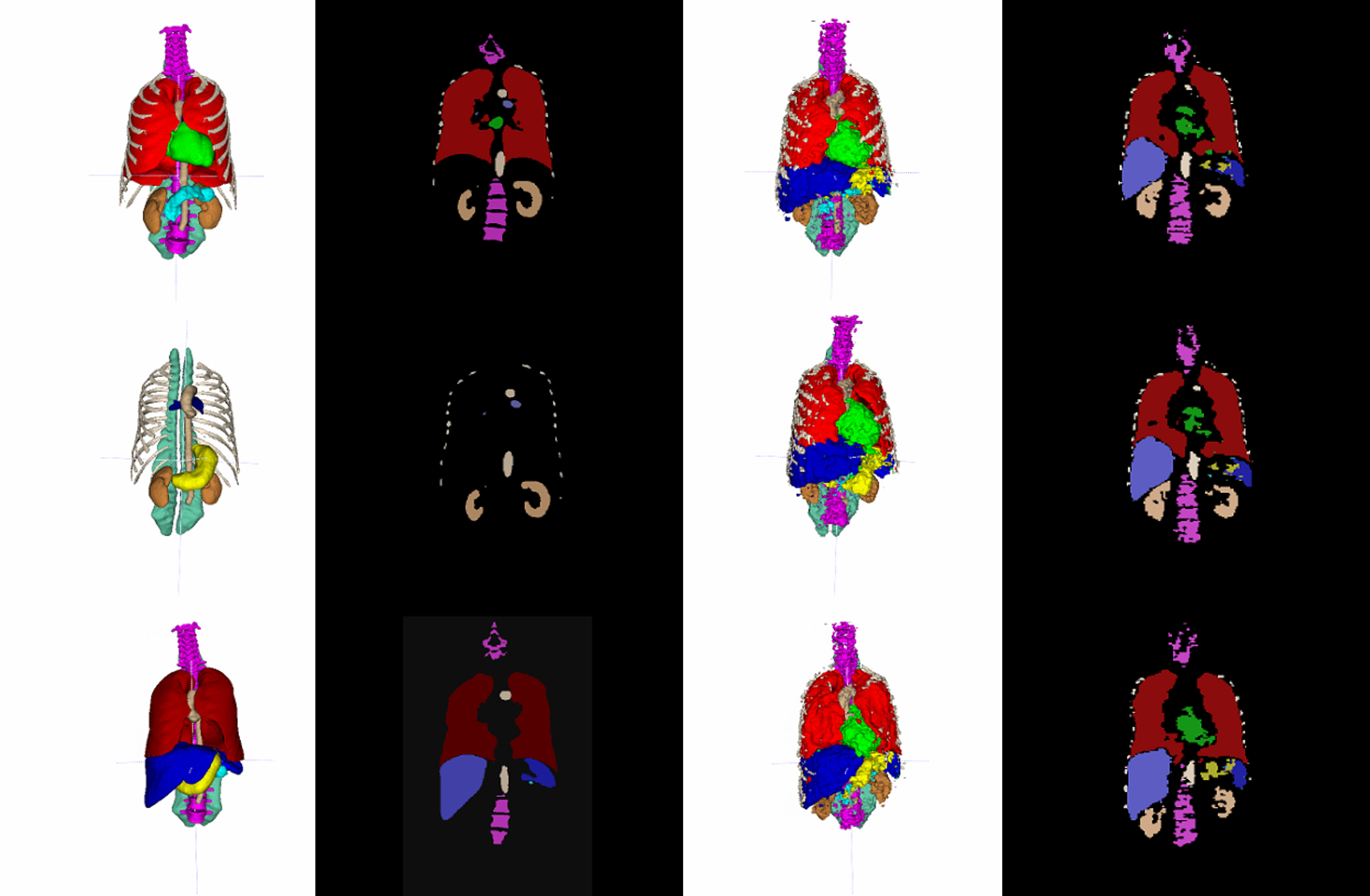}
 \caption{Qualitative results of multi-class anatomy completion. The first and second column show three incomplete instances from the same subject in 3D and coronal views. The last two columns show the corresponding reconstruction results.}
\label{multi_class_results}
\end{figure}

For the multi-class experiment, we choose 12 anatomies, including the lung, heart, spleen, stomach, pancreas, spine, rib cage, liver, kidney, aorta, a pair of autochthon muscles, and the pulmonary artery (Figure \ref{multi_class} (A)). We extract the 12 above-mentioned anatomy segmentations from 18 whole-body segmentations randomly chosen from the training set. We create 10 incomplete instances for each case by randomly removing some of the 12 anatomies (e.g., Figure \ref{multi_class} (B-D)), resulting in $18 \times 10 =180$ training samples. Images are resized to $256 \times 256 \times 128$ ($L,W =256, H=128$) and the $DAE_{agg}$ method is used for the experiment (i.e., to learn a \textit{many-to-one mapping}). Figure \ref{multi_class_results} presents the multi-class anatomy completion results on test samples that are not involved during training. It is noticeable from the reconstructions that the long thin structures i.e., the ribs, are not well reconstructed (e.g., the last row of Figure \ref{multi_class_results}). Terracing artifacts are also obvious on the reconstructed anatomical shapes compared to the ground truth, which can be partly attributed to downsampling.

\section{Discussion and Conclusion}
In this paper, we demonstrated that multi-class anatomy reconstruction can be realized in a single shape completion framework. Given an incomplete instance with random missing anatomies, a DAE network reconstructs the missing anatomies specific to the instance, so that the new reconstructions geometrically align with existing anatomies. We further verified that residual learning and loss aggregation can significantly boost the performance of the DAE for the reconstructive task, and mitigate the low sensitivity and false prediction issues. Besides the baseline DAE, residual connection and loss aggregation can be easily implemented on top of more complicated network architectures. The models can not only reconstruct multiple missing anatomies simultaneously (Figure \ref{completion_results}) but also a specific anatomy, despite their sizes (Figure \ref{anatomy_results} and Appendix A). There are several known limitations remaining to be addressed in future work: (i) Not all anatomy classes are covered by the segmentations of the CT dataset, such as the skull, full limb, brain, skins and soft tissues (e.g., facial soft tissues and most of the muscles); (ii) A quantitative evaluation for each specific anatomy is lacking (only qualitative results are provided in Figure \ref{anatomy_results} and Appendix A); (iii) The reconstructions from the multi-class anatomy completor suffer from terracing artifacts and discontinuous ribs. A super-resolution procedure can be applied to refine the initial reconstructions using sparse convolutional neural networks \cite{kroviakov2021sparse}. An interesting direction for future work is to use the multi-class anatomy completor in whole-body segmentation, where it can be used to generate the initial pseudo labels of the organs given only skeletal annotations (e.g., the rib case and spine. See Appendix B).

\section*{Acknowledgement}
The work is supported by the Plattform für KI-Translation Essen (KITE) from the REACT-EU initiative (EFRE-0801977, \url{https://kite.ikim.nrw/}) and „NUM 2.0“ (FKZ: 01KX2121). The anatomical shape dataset used in this paper can be accessed through \textit{MedShapeNet} at \url{https://medshapenet.ikim.nrw/}.

\newpage

\begin{subappendices}
\section*{Appendix A. Reconstructing Small Anatomies}

\begin{figure}[h!]
\renewcommand\thefigure{A}
\centering
\includegraphics[width=1\linewidth]{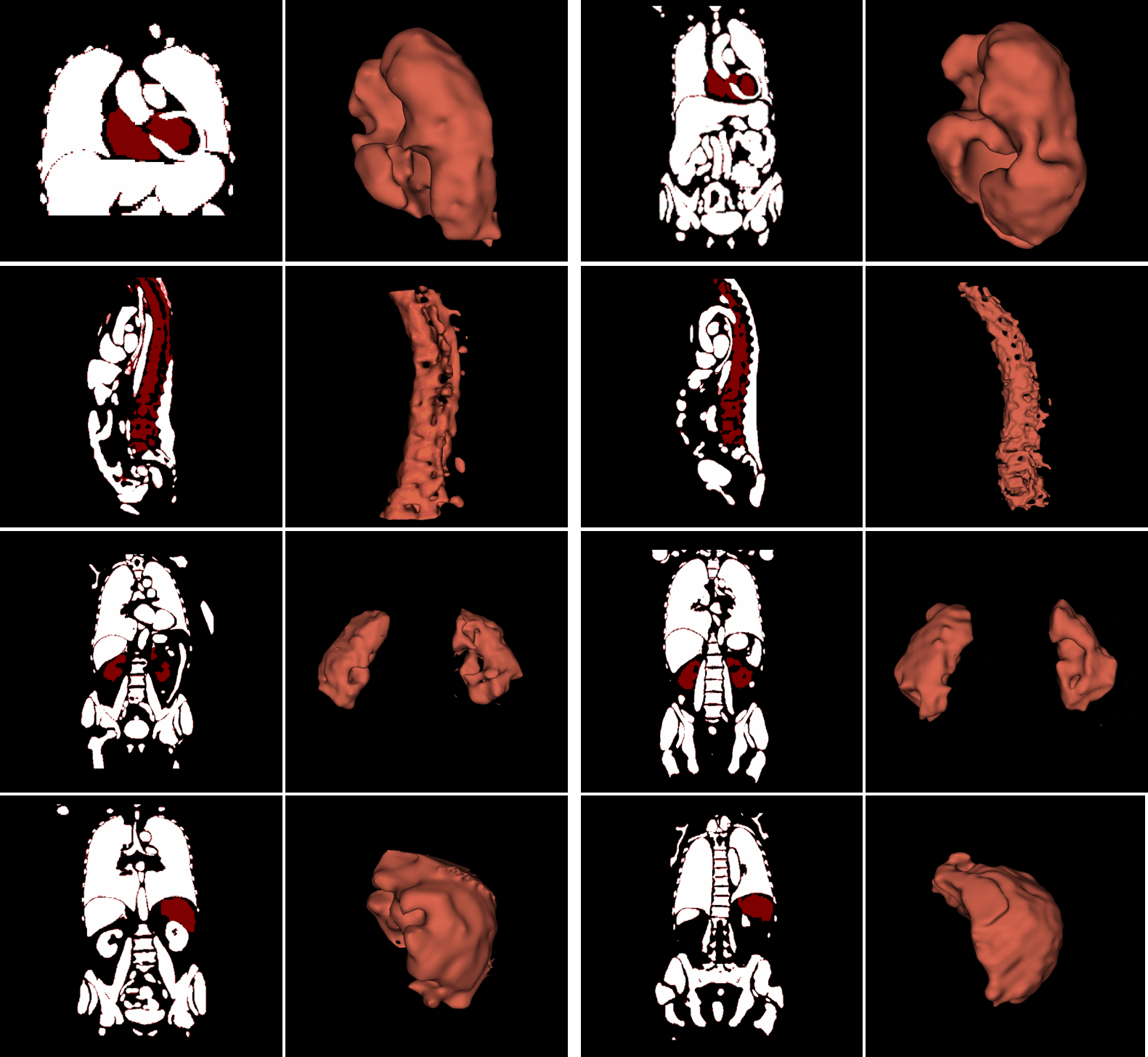} 
 \caption{Reconstruction results of individual, small anatomies by $DAE_{agg+res}$ trained with an increased loss aggregation scope ($M$). From the top: heart (2.4\%), spine (4.3\%), kidney (1.7\%) and spleen (1.2\%). The percentages in the brackets are the approximate volume ratio of the anatomy to the corresponding whole-body segmentations. The preliminary results demonstrate that increasing $M$ (in Equation 3, 4 in the main manuscript) increases also the sensitivity of the reconstructive model, which helps the model identify and reconstruct very small anatomies. Two test instances are presented for each anatomy class.}
\label{supplementary}
\end{figure}

\section*{Appendix B. Anatomy Completion from Skeletons (rib cage + spine)}
\label{appendixb}
\begin{figure}[h!]
\renewcommand\thefigure{B}
\centering
\includegraphics[width=1\linewidth]{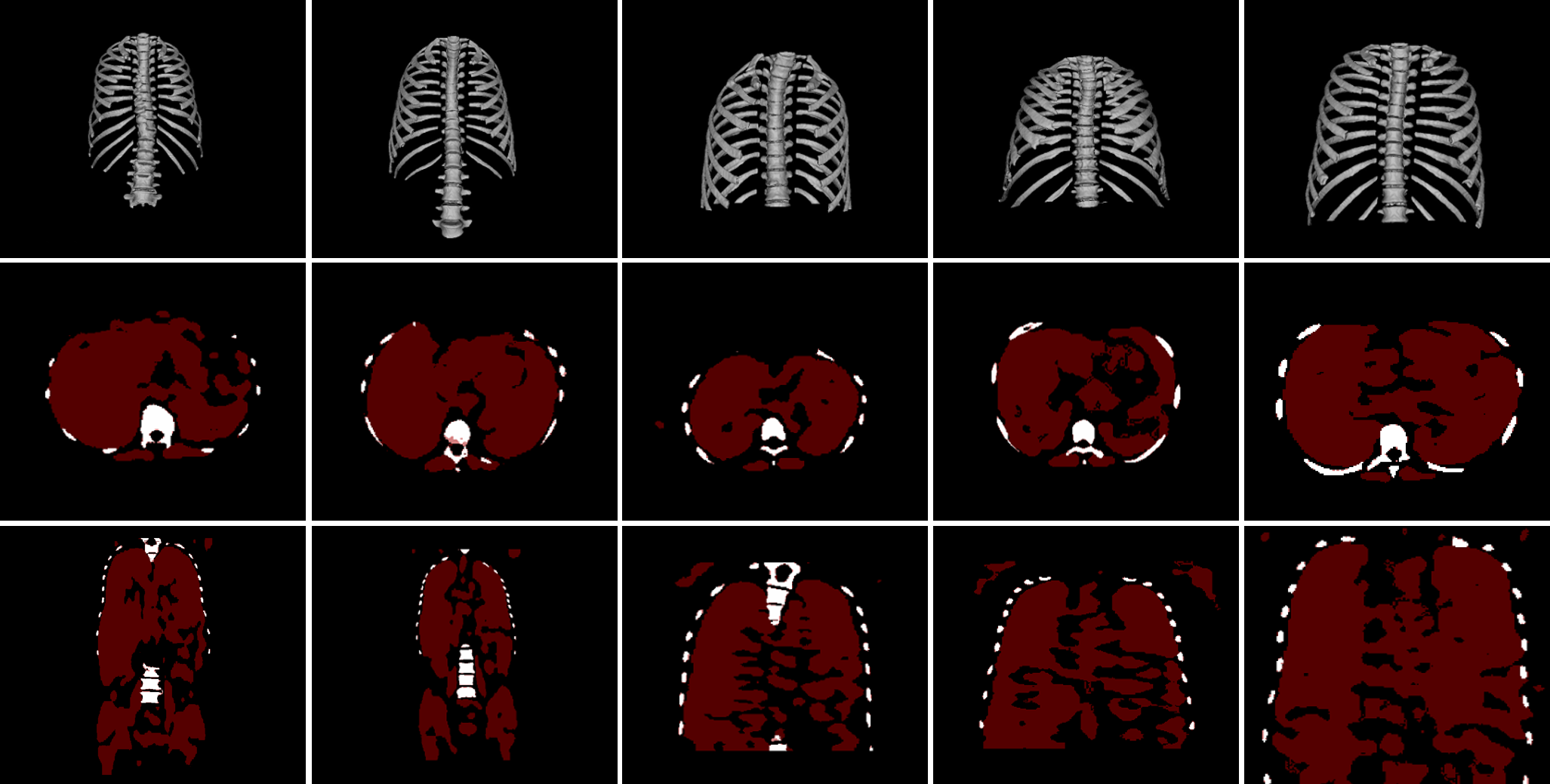} 
 \caption{The first row shows the input skeleton (ribs and spine), and the second to third row show the reconstruction results in axial and coronal views, respectively. The results are obtained by training $DAE_{res}$ on 40 such 'skeleton-full' pairs for 200 epochs. The preliminary results demonstrate the feasibility of reconstructing the full anatomy based only on the skeleton.}
\label{supplementary1}
\end{figure}

\end{subappendices}

\newpage
\bibliographystyle{splncs04}
\bibliography{mybibliography}

\end{document}